\documentclass[12pt,preprint]{aastex}

\slugcomment{Accepted by the Astrophysical Journal 15 Dec 2010}

\shorttitle{Self-Consistent Non-Linear Modeling of Solar Active Regions}

\shortauthors{Wheatland and Leka}

\begin{document}

\title{Achieving Self-Consistent Nonlinear Force-free Modeling of Solar 
  Active Regions}

\author{M.~S. Wheatland}
\affil{Sydney Institute for Astronomy, School of Physics, 
  The University of Sydney, NSW 2006, Australia}
\email{michael.wheatland@sydney.edu.au}

\and

\author{K.~D. Leka}
\affil{North West Research Associates, 
  Colorado Research Associates Division, 
  3380 Mitchell Lane, Boulder, CO 80301, USA}
\email{leka@cora.nwra.com}

\begin{abstract}
A nonlinear force-free solution is constructed for the coronal magnetic
field in NOAA solar active region AR 10953 based on a photospheric 
vector magnetogram derived from Hinode satellite observations on 30 
April 2007, taking into account uncertainties in the boundary data
and using improved methods for merging multiple-instrument data. 
The solution demonstrates the `self-consistency' procedure of Wheatland 
\& R\'{e}gnier (2009), for the first time including uncertainties. 
The self-consistency procedure addresses the problem that photospheric 
vector magnetogram data are inconsistent with the force-free model, and 
in particular that the boundary conditions on vertical electric current 
density are over-specified and permit the construction of two different
nonlinear force-free solutions. The procedure modifies the boundary 
conditions on current density during a sequence of cycles until the 
two nonlinear force-free solutions agree. It hence constructs an 
accurate single solution to the force-free model, with boundary 
values close, but not matched exactly, to the vector magnetogram 
data. The inclusion of uncertainties preserves the boundary conditions 
more closely at points with smaller uncertainties. The self-consistent 
solution obtained for active region AR 10953 is significantly 
non-potential, with magnetic energy $E/E_0\approx 1.08$, where $E_0$ is 
the energy of the reference potential (current-free) magnetic field. 
The self-consistent solution is shown to be robust against changes in 
the details of the construction of the two force-free models at each 
cycle. This suggests that reliable nonlinear force-free modeling of 
active regions is possible if uncertainties in vector magnetogram 
boundary data are included.
\end{abstract}


\section{Introduction\label{sec1}}

Intense magnetic fields around sunspot regions drive the most energetic
examples of solar activity -- large solar flares and coronal mass
ejections -- which can produce hazardous space weather conditions close
to Earth.
Severe space weather storms present a variety of hazards including
increased radiation levels for space travellers and crew on polar 
air flights, the risks of disabling
of GPS systems and damage to power grids \citep{2008sswe.rept.....C},
and potential large economic losses due to damage to
communications satellites (Odenwald et al.\ 2006). These effects
motivate a need to understand and model the source regions of space
weather threats at the Sun.

Spectro-polarimetric measurements of magnetically sensitive photospheric
lines may be used to infer the vector magnetic field at the surface of
the Sun \citep{LanLan04}, and the resulting maps of the field are called 
vector magnetograms. A new generation of instrumentation is 
set to provide vector magnetogram data over the next solar cycle with
unprecedented quality, resolution, and temporal cadence, including the
Solar Optical Telescope Spectro-Polarimeter (SOT/SP) on the Hinode
spacecraft (Tsuneta et al.\ 2008) and the Helioseismic and Magnetic
Imager on the Solar Dynamics Observatory (SDO/HMI) 
\citep{2006cosp...36.1469S,2007SoPh..240..177B}. In principle these 
measurements provide boundary values for coronal 
magnetic field modeling, or `magnetic field reconstruction'. There is a 
pressing need for this modeling capability, which has many potential 
applications in solar research. However, studies have revealed basic 
difficulties preventing the construction of reliable coronal field 
models from vector magnetograms using present modeling techniques. 
Specifically, a popular and well-developed approach to magnetic field 
reconstruction, nonlinear force-free modeling, has been shown to 
produce inconsistent results when applied to Hinode/SOT vector 
magnetogram data 
\citep{2008ApJ...675.1637S,2009ApJ...696.1780D,WheEtAl2010}. Different
modeling codes produce different results based on the same data, and
the results from individual codes are also internally inconsistent.
As described in more detail in section~\ref{sec2}, vector magnetograms
allow for two sets of boundary conditions for the force-free model,
and the two solutions do not in general agree. It is also difficult
to construct the solutions: the boundary conditions on electric current
present large currents which prevent the force-free methods from
converging strictly. As a result the `solutions' may be inaccurate
with respect to the model. The modeling has been shown to be unable to 
provide a reliable estimate of the free magnetic energy of a solar active 
region. Further details of the problems are provided in 
section~\ref{sec2}.

One likely cause of the failure of nonlinear force-free modeling is
a basic inconsistency between the boundary data and the model: 
the model includes only magnetic forces, an assumption 
that may be warranted in the corona, but which is thought not to apply 
at the photospheric level where the data originate 
\citep{1995ApJ...439..474M}.
Recently a new approach to coronal magnetic field reconstruction was
presented to address this problem \citep{2009ApJ...700L..88W}. The
`self-consistency' procedure (described in more detail in
section~\ref{sec2}) identifies a nonlinear force-free solution with
boundary values close to those implied by the vector magnetogram data. 
The method 
constructs a single solution to the nonlinear force-free model which 
strictly converges. The solution does not match exactly the field 
at the photospheric level, but may describe the coronal
magnetic field. In \citet{2009ApJ...700L..88W} the method was 
demonstrated in application to Hinode/SOT
data for NOAA active region 10953. A solution to the model was
constructed with energy $E/E_0\approx 1.02$, where $E_0$ is the energy
of the reference potential (current-free) magnetic field matching the 
boundary conditions on the vertical component of the field. The 
self-consistency procedure is capable of taking into account 
relative uncertainties in
the solar boundary data, so that the boundary values for the solution
match the vector magnetogram data more closely at points with small
uncertainties. However, uncertainties were omitted from the
\citet{2009ApJ...700L..88W} calculations, and this omission accounts in
part for the low value of the magnetic free energy of the constructed
solution ($2\%$ of the energy of the potential field), as explained
in section~\ref{sec24}. For this reason
the results in \citet{2009ApJ...700L..88W} were argued to represent a
`proof in concept' of the self-consistency procedure, rather than a 
full demonstration of 
its modeling capability. In this paper we return to the problem of 
modeling AR 10953, including uncertainties in the calculation.

The detailed presentation of this paper is as follows. 
Section~\ref{sec2}
provides a more detailed account of the background to the modeling
presented here, incuding a summary of the nonlinear force-free model
(section~\ref{sec21}), a description of the failure of the model in 
application to solar data (section~\ref{sec22}), and an account of the 
self-consistency procedure (section~\ref{sec23}) and its initial 
application to AR 10953 by \citet{2009ApJ...700L..88W} 
(section~\ref{sec24}). A reader familiar with the background may skim 
or omit section~\ref{sec2}, but it provides necessary background for 
a reader new to the topic. Section~\ref{sec3} describes the  
modeling of AR 10953 taking into account uncertainties, with 
section~\ref{sec31} presenting the data, and section~\ref{sec32} the 
results of the modeling. Section~\ref{sec4} discusses the results.
 
\section{Background\label{sec2}}

\subsection{Nonlinear force-free model\label{sec21}}

Magnetic fields in the corona are often described as being force free
\citep{1994plph.book.....S}. The force-free model is static and 
includes only the magnetic (Lorentz) force, an approximation justified 
by the low ratio of gas pressure to magnetic pressure likely to exist 
at most locations in the corona 
\citep{1995ApJ...439..474M,2001SoPh..203...71G}. A force-free magnetic 
field ${\bf B}$ satisfies 
\begin{equation}\label{eq:force_free}
{\bf B}\cdot \nabla \alpha =0 
\quad \mbox{and} \quad
\nabla \times {\bf B} =\alpha {\bf B},
\end{equation}
where $\alpha$ is the force-free parameter. This parameter is constant 
along field lines, but varies (in general) from field line 
to field line, and the implicit dependence of $\alpha$ on ${\bf B}$
makes the model nonlinear. The boundary conditions required by
the model are the normal component of ${\bf B}$ in the boundary 
(denoted $B_n$), and a specification of $\alpha $ over one polarity 
(sign) of $B_n$~(e.g.\ Grad \& Rubin 1958; Sakurai 1981; Aly 1989; 
Amari et al.\ 2006). We denote the two polarities $P$ (positive) and
$N$ (negative), for boundary points with $B_n>0$ and $B_n<0$ 
respectively (assuming a normal directed into the coronal volume). 
The boundary condition on $\alpha$ is equivalent to a 
specification of the normal component of the electric current density 
${\bf J}$ over either $P$ or $N$, because ${\bf J}=\alpha {\bf B}/\mu_0$ 
in the model.

A number of numerical methods of solution of the nonlinear force-free 
equations have been developed in the solar context [for recent
reviews see e.g.\
\citet{2005ESASP.596E..12N,2007GApFD.101..249A,2007MmSAI..78..126R,
2008JGRA..11303S02W}].
Most methods focus on the solution of the problem in a 
half space ($z>0$), with the lower boundary ($z=0$) representing the 
photosphere, solar curvature being neglected. The different methods 
vary in terms of numerical methods and in the treatment of boundary 
conditions. Some methods -- e.g.\ the optimization method 
\citep{2000ApJ...540.1150W,2004SoPh..219...87W} -- use the full vector 
magnetic field over both $P$ and $N$ in the lower boundary as a 
boundary condition. This is formally an over-prescription, but does 
not introduce a problem provided the boundary conditions are consistent
with the model, as shown by application to test cases 
\citep{2006SoPh..235..161S}. Methods based on the
`current-field iteration' or Grad-Rubin procedure
\citep{GradRubin58,1981SoPh...69..343S,1997SoPh..174..129A,2006A&A...446..691A}
use the formally correct
boundary conditions on $\alpha$ described above. A Grad-Rubin
implementation is the basis for the calculations 
in this paper \citep{2007SoPh..245..251W}. These methods work by
replacing the nonlinear force-free equations by a sequence of 
iterations, such that at the $i^{\rm th}$ iteration the linear 
equations 
\begin{equation}\label{eq:grad_rubin}
{\bf B}^{[i-1]}\cdot \nabla \alpha^{[i]} =0
\quad \mbox{and} \quad
\nabla \times {\bf B}^{[i]} =\alpha^{[i]} {\bf B}^{[i-1]},
\end{equation}
are solved for ${\bf B}^{[i]}$ given ${\bf B}^{[i-1]}$, subject to 
$\nabla\cdot {\bf B}^{[i]}=0$ and the boundary conditions on
$\alpha^{[i]}$ and ${\bf B}^{[i]}$. For the half space problem
the boundary conditions are the specification of $B_z^{[i]}$ at 
$z=0$ and $\alpha^{[i]}$ at $z=0$ over either $P$ or $N$. The 
iteration is usually started by setting ${\bf B}^{[0]}$ equal to the 
reference potential (current-free) field matching the boundary 
conditions on $B_z$. A fixed point of the iteration scheme
specified by Equations~(\ref{eq:grad_rubin}) provides a solution to the
nonlinear force-free Equations~(\ref{eq:force_free}).

\subsection{Failure of nonlinear force-free modeling for solar data
\label{sec22}}

Although the nonlinear force-free methods work well in application to
test cases \citep{2006SoPh..235..161S} and to synthetic solar-like data
\citep{2008SoPh..247..269M}, the application to vector magnetogram data
has revealed serious problems. Tests of different 
methods using Hinode/SOT data for NOAA active regions AR 10930
\citep{2008ApJ...675.1637S} and AR 10953 \citep{2009ApJ...696.1780D}
produced very different solutions for each active region depending on
the specific method used. In particular the estimated magnetic free
energy of each active region varied substantially among the solutions,
preventing reliable determination of this quantity. Perhaps worse, the
solutions from particular methods were also found to be internally
inconsistent. For example, for the Grad-Rubin methods the vector
magnetogram data provide both the $P$ and $N$ choices for the boundary
conditions on the force-free parameter. Specifically, $\alpha$ may be
estimated from the data at the photosphere ($z=0$) by calculating the
gradients in the observationally-inferred transverse fields $B_x$ and
$B_y$ according to
\begin{equation}\label{eq:vmag_alpha}
\alpha 
  =\frac{1}{B_z}\left(\frac{\partial B_y}{\partial x}
  -\frac{\partial B_x}{\partial y}\right),
\end{equation}
and these values are available over both $P$ and $N$. The $P$ and $N$
solutions obtained by applying the Wheatland (2007) current-field
iteration method to the data for active regions AR 10930 and AR 10953 
were found to be very different 
\citep{2008ApJ...675.1637S,2009ApJ...696.1780D}.

It is likely that a number of factors contribute to the failure of the
modeling. Vector magnetogram field values are approximate
due to resolution limitations, modeling assumptions, and photon noise 
in the polarization measurements.
Assumptions made in the inference of the field from the
Stokes spectral measurements \citep{1996SoPh..164..169D,LanLan04}, 
for example using a Milne-Eddington atmosphere, introduce 
inaccuracy~\citep{1977SoPh...55...47A,1987ApJ...322..473S,2007A&A...464..323B}.
The values are also uncertain due to the influence of instrumental 
noise and resolution limitations on the polarization measurements.
The component of the magnetic field transverse to the line of sight 
is particularly uncertain due to the intrinsically smaller linear 
polarization signal relative to the integrated intensity, 
magneto-optical effects \citep{landolfilandi82}, and the need to 
resolve an intrinsic 180-degree ambiguity in the azimuthal direction of the 
field 
\citep{1994SoPh..155..235M,2006SoPh..237..267M,2009SoPh..260..271C}. 

A problem with the force-free modeling is that different methods of
solving the Equations~(\ref{eq:force_free}) make different 
assumptions regarding the problem of the `missing' boundary 
conditions at the sides and top boundaries of the computational 
domain.  These different assumptions lead (not surprisingly) to 
different results \citep{2006SoPh..235..161S,2008SoPh..247..269M,
2008ApJ...675.1637S,2009ApJ...696.1780D}. However, the
modeling also faces a more fundamental difficulty: the magnetic 
field at the height of the measurements is unlikely to be force free 
because the dense photospheric gas is subject to pressure, gravity, 
and dynamical forces in addition to the magnetic forces 
\citep{1995ApJ...439..474M}. The boundary data are fundamentally 
inconsistent with the model.

The accuracy of the force-free assumption for the boundary data may be
checked a priori by calculating integrals of the field in the boundary
representing the net magnetic flux, the net force and the torque on the
field in $z>0$
\citep{1969SvA....12..585M,1984ApJ...283..349A,1989SoPh..120...19A}. The
requirement that these integrals vanish provides a set of necessary
conditions for the existence of a force-free solution for given boundary
data. In general the integrals are found to be substantially different
from zero for photospheric vector magnetogram data. One treatment of 
this problem is a procedure called `preprocessing,' whereby the
boundary data are varied to minimize the integrals, subject to
constraints \citep{2006SoPh..233..215W,2008SoPh..247..249W}. 
Preprocessing has been found to improve nonlinear 
force-free reconstructions from synthetic solar data for some methods 
\citep{2008SoPh..247..269M}, and preprocessed data were used in the 
modeling of active regions AR 10930 and AR 10953 presented by 
\citet{2008ApJ...675.1637S} and \citet{2009ApJ...696.1780D}
respectively.
However, there are non-trivial problems with the preprocessing 
approach. First, the conditions being minimized are necessary but not 
sufficient, so there is an arbitrary number of such constraints
\citep{1989SoPh..120...19A,2007MmSAI..78..126R}. The minimization
procedure will not satisfy all such constraints, and it is
straightforward to show that preprocessed boundary data are still
inconsistent with the force-free model \citep{2009ApJ...696.1780D}. As 
a result, the $P$ and $N$ solutions constructed by Grad-Rubin methods 
still disagree, and `force-free solutions' constructed using other 
methods, e.g.\ the optimization method 
\citep{2000ApJ...540.1150W,2004SoPh..219...87W} must be internally 
inconsistent with the force-free model. The situation is then that the
observed boundary data is not matched by the `solution,' 
and the force-free model is not being accurately solved. A second 
problem is that preprocessing typically involves a smoothing of the 
data, which is undesirable because it discards information. Recently, 
another solution was proposed, in which a vector field is calculated
which attempts to simultaneously minimize the Lorentz force, the
divergence, and the departure from observed boundary conditions, using
an optimization-like procedure \citep{2010A&A...516A.107W}. However,
this method also fails to adhere to the force-free equations or
retain (to within measurement uncertainties) the observed boundary 
data.

A related problem with solar nonlinear force-free modeling is that 
the boundary conditions on the vertical electric current density 
$J_z$ corresponding to vector-magnetogram derived values of 
$\alpha$ include very large electric currents. The values of the
current density are considerably uncertain because of the uncertainties
in the field determinations discussed above, and there is also
dependence of the values on the observational spatial resolution,
which changes when field values are interpolated onto new grids 
and rebinned \citep{1999SoPh..188...21L,2009SoPh..260..271C}.
Nevertheless there appear to be electric currents sufficiently large
to be inconsistent with the force-free model, which prevent strict 
convergence of nonlinear force-free methods 
\citep{2008ApJ...675.1637S,2009ApJ...696.1780D}. The
\citet{2007SoPh..245..251W} method is often found to reach a state 
where the field configuration exhibits slow oscillation as a function 
of Grad-Rubin iteration number rather than achieving a fixed 
state. The energy of the solution also oscillates slowly as a 
function of iteration number. This behavior suggests the non-existence, 
or the non-uniqueness, of a solution to the boundary value problem 
being solved \citep{2006A&A...446..691A,2006SoPh..235..201I}. 
The consequence of this problem for the modeling is that the 
solution is non-unique (the exact result depends on the arbitrary 
choice of a stopping iteration) and also the force-free model is
being solved less accurately (the field is less force free than in 
the case of strict convergence because the Grad-Rubin procedure does
not achieve a fixed point). 

\subsection{Self-consistency procedure\label{sec23}}

In this paper we adopt the approach to resolving the inconsistency 
problem introduced
recently by \citet{2009ApJ...700L..88W}. The `self-consistency'
procedure identifies one nonlinear force-free solution with 
boundary values close, but not exactly matched, to the vector 
magnetogram data, with guidance from the estimated uncertainties.
The result is an accurate solution within the assumptions of the 
force-free model. The procedure may be summarized as follows.

\begin{figure}
\plotone{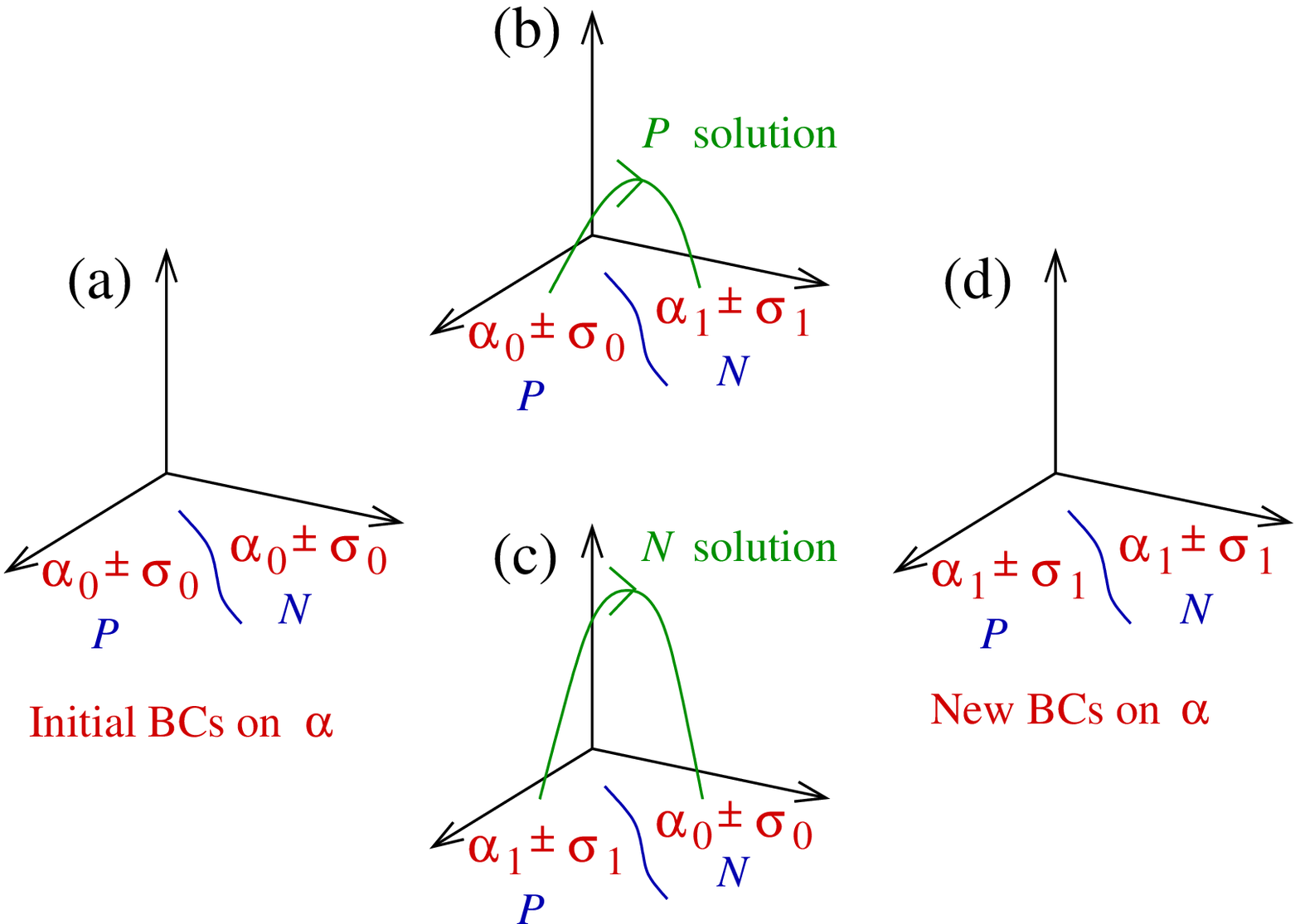}
\caption{The first step of the self-consistency procedure. 
Two solutions are constructed from the $P$ (positive polarity) and 
$N$ (negative polarity) boundary values of the force-free parameter 
$\alpha_0$, shown in panel (a). The $P$ and $N$ solutions map 
values of $\alpha_0\pm\sigma_0$ along magnetic field lines from 
$P\rightarrow N$ and from $N\rightarrow P$ respectively, as 
illustrated in panels (b) and (c) respectively. These mappings define
a new set of boundary conditions ($\alpha_1\pm\sigma_1$) over $P$ and 
$N$, as shown in panel (d). The blue curves in the boundary in each
panel indicate the neutral line dividing $P$ and $N$, and the green 
curves in panels (b) and (c) are field lines of the $P$ and $N$ 
solutions, with arrows indicating field direction.
\label{fig:SCmethod}}
\end{figure}

First, $P$ and $N$ nonlinear force-free solutions are constructed,
from unpreprocessed vector magnetogram boundary data, using the
Grad-Rubin code of \citet{2007SoPh..245..251W}, with a chosen number 
of iterations. These
solutions use boundary values $\alpha_0\pm\sigma_0$ for the force-free
parameter, where the $\alpha_0$ values are derived from the vector
magnetogram data via Equation~(\ref{eq:vmag_alpha}), and the
uncertainties $\sigma_0$ are obtained by propagating the uncertainties 
assigned to the field values~\citep{1999SoPh..188....3L}. The $P$ 
solution provides a mapping, along field lines, of the observational 
boundary values $\alpha_0\pm\sigma_0$ at points in the region $P$ to 
points in $N$. This defines new possible boundary values 
$\alpha_1\pm\sigma_1$ of the force-free parameter at points in $N$. 
(These values will not match the observational values 
$\alpha_0\pm\sigma_0$ due to the inconsistency problem.) Similarly, 
the $N$ solution maps the observational boundary values 
$\alpha_0\pm\sigma_0$ at points $N$ to points in $P$, and so defines 
new possible boundary values $\alpha_1\pm\sigma_1$ at points in $P$. 
Together the two solutions define a complete new set of boundary
values $\alpha_1\pm\sigma_1$ of the force-free parameter (over $P$ and
$N$). Figure~\ref{fig:SCmethod} illustrates the first step of the
self-consistency procedure.

Second, Bayesian probability theory \citep{2003prth.book.....J} is 
applied to decide on a boundary value of the force-free parameter at 
each boundary point, based on the two sets of values 
$\alpha_0\pm\sigma_0$ and $\alpha_1\pm\sigma_1$. The idea is that 
the values represent two observations subject to Gaussian 
uncertainties, and Bayes's theorem is applied to determine the true 
value of the force-free parameter. The resulting most probable value 
and associated uncertainty are given by the uncertainty-weighted 
averages \citep{2009ApJ...700L..88W}:
\begin{equation}\label{eq:bayes_avg}
\alpha_2 = \frac{
  \alpha_0/\sigma_0^2+\alpha_1/\sigma_1^2}
              {1/\sigma_0^2+1/\sigma_1^2}
\qquad \mbox{and} \qquad
\sigma_2=\left(\frac{1}{\sigma_0^2}+\frac{1}{\sigma_1^2}\right)^{-1/2}.
\end{equation}
The resulting set of boundary values $\alpha_2\pm \sigma_2$ over 
$P$ and $N$ is still, in general, inconsistent with the force-free 
model, but the values are expected to be closer to consistency.

Finally, steps one and two are iterated. Two force-free solutions 
(the $P$ and $N$ solutions) are constructed from the new boundary values
$\alpha_2\pm\sigma_2$, again by applying current-field 
iteration, in each case. The two solutions define new mappings
between the $P$ and $N$ regions, and hence a complete new set of
boundary conditions $\alpha_3\pm\sigma_3$. 
Equations~(\ref{eq:bayes_avg}) are applied to the two sets of values
$\alpha_2\pm\sigma_2$ and $\alpha_3\pm\sigma_3$ to decide on most 
probable values $\alpha_4\pm\sigma_4$ at each boundary point, and so 
on. Each iteration of the self-consistency procedure [i.e.\ the 
construction of the $P$ and $N$ solutions, and then the application of 
Equations~(\ref{eq:bayes_avg}) to give a new set of boundary conditions
on current] is referred to as a `self-consistency cycle.' The cycles
are numbered starting at one, and the $P$ and $N$ solutions are 
enumerated with an index $k$, such that the first cycle involves 
construction of $P$ and $N$ solutions for the vector magnetogram
boundary values $\alpha_0\pm \sigma_0$, with the solutions
numbered $k=1$ and $k=2$ respectively. 

Self-consistency cycling is 
expected to converge to a self-consistent solution, i.e.\ a set of 
boundary values of the force-free parameter for which the $P$ and $N$ 
solutions are the same. The result is a single force-free solution, 
with (in some sense) a minimal departure from the observations.

\subsection{Wheatland \& R\'{e}gnier (2009)\label{sec24}}

In \citet{2009ApJ...700L..88W} the self-consistency procedure was 
demonstrated in 
application to Hinode/SOT data for AR 10953 observed at 22:30~UT 
on 30 April 2007. The vector magnetogram was the same as 
that described in \citet{2009ApJ...696.1780D} but without the 
preprocessing (and associated smoothing). 
Uncertainties were assumed to be equal at each boundary point. In that 
case the first of Equations~(\ref{eq:bayes_avg}) reduces to 
$\alpha_2 = \frac{1}{2} \left(\alpha_0+\alpha_1\right)$, i.e.\ a 
simple average of the two values. The self-consistency procedure was
applied for ten cycles and found to converge to a solution with an
energy $E/E_0\approx 1.02$, where $E_0$ is the energy of the reference 
potential field.

\begin{figure}
\plotone{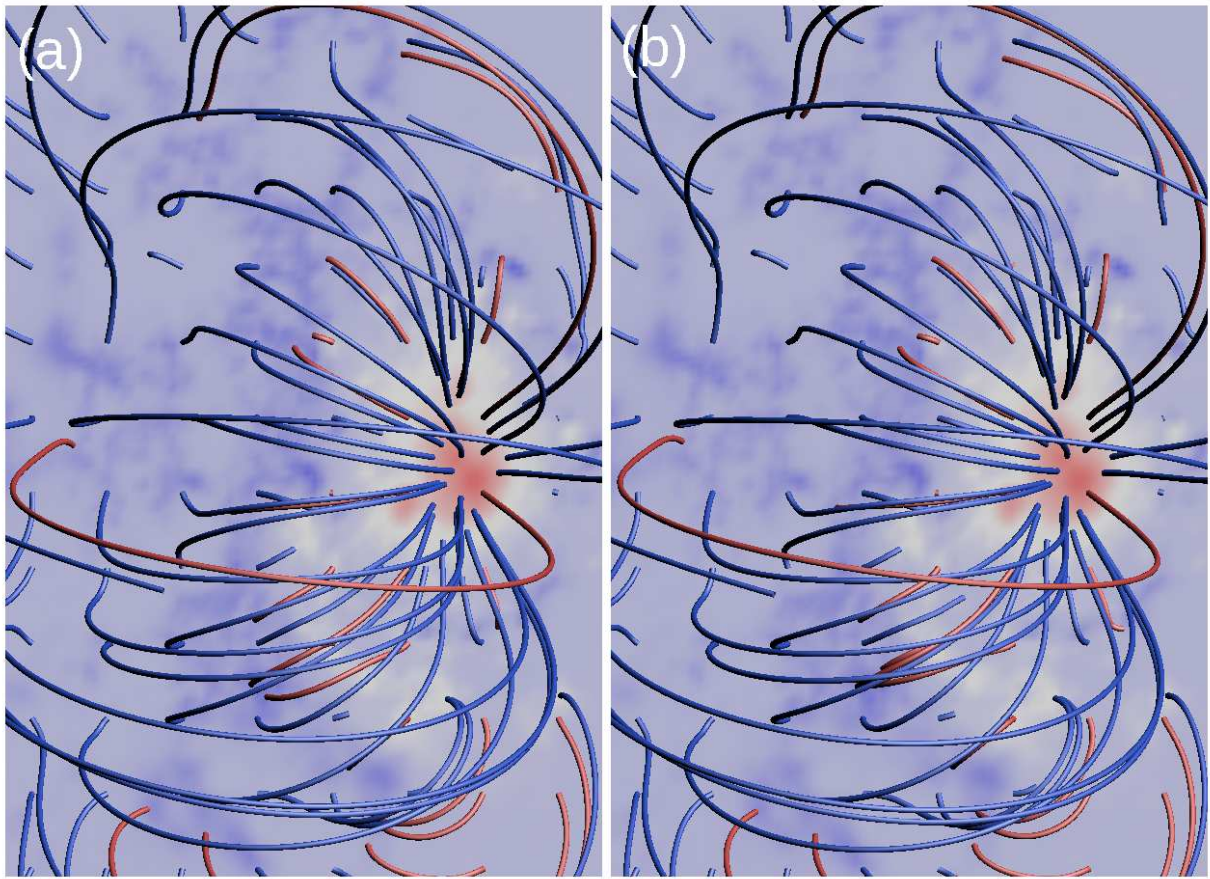}
\caption{The self-consistent solution for the coronal magnetic field
of NOAA AR 10953 obtained by \citet{2009ApJ...700L..88W}. Panel (a) 
shows the $P$ solution and panel (b) shows the $N$ solution,
after 10 self-consistency cycles. The color image in the background in
each case shows the boundary values of $B_z$, with negative values red, 
values close to zero white, and positive values blue. The blue curves 
in each panel are field lines originating at points in the $P$ 
polarity, and the red curves are field lines originating at points in
the $N$ polarity. The sets of field lines for the $P$ solution in 
panel (a) and for the $N$ solution in panel (b) agree closely.
\label{fig:WheReg09Stereo}}
\end{figure}

Figure~\ref{fig:WheReg09Stereo} illustrates the self-consistent 
solution obtained by \citet{2009ApJ...700L..88W}. Panel (a) shows
the $P$ solution and panel (b) shows the $N$ solution, after
10 self-consistency cycles. Each image looks down on the central part 
of the computational volume. The red-white-blue image in the 
background shows the boundary values of $B_z$, with red indicating 
negative values, white values close to zero, and blue positive values. 
Two sets of field lines are shown in each panel by the red and the blue 
curves. The blue curves are field lines originating at points in the
$P$ polarity and the red curves are field lines originating at points
in the $N$ polarity. The sets of field lines in panel (a) and
in panel (b) agree closely, illustrating the achievement of 
self-consistency. Figure~1 in \citet{2009ApJ...700L..88W} (not 
reproduced here) also shows the $P$ and $N$ solutions at the 
first self-consistency cycle, i.e.\ calculated directly from the 
original boundary data. The initial $P$ and $N$ solutions disagree 
markedly, with the $N$ solution having more highly distorted magnetic 
field lines than the $P$ solution, due to large electric currents 
associated with the leading polarity negative spot in the region.

The self-consistent solution shown in Figure~\ref{fig:WheReg09Stereo} has 
a relatively small magnetic free energy (about $2\%$ of the energy of 
$E_0$). As discussed in \citet{2009ApJ...700L..88W}, this is due in 
part to the neglect of uncertainties in the boundary values of $\alpha$. 
Many of the boundary points have zero values of $\alpha$. The Hinode 
field of view is small, and to compensate for this the Hinode data was
embedded in a larger field of view based on a line-of-sight magnetogram 
from the Michelson Doppler Interferometer (MDI) instrument on the Solar 
and Heliospheric Observatory (SOHO) spacecraft 
\citep{1995SoPh..162..129S}. The SOHO/MDI data points have zero values 
for $J_z$ and hence for $\alpha$, and these zero boundary 
values are given equal weight to non-zero Hinode-derived values of
$\alpha$ in the self-consistency averaging of
the first of Equations~(\ref{eq:bayes_avg}). This leads to a reduction
in the electric currents in the solution, and hence a reduction in
the magnetic free energy. Also, some strong currents are expected to
occur in strong field regions, which will have more 
accurately-determined field values in the vector magnetograms (smaller 
uncertainties). These larger currents would be preserved in the
application of Equations~(\ref{eq:bayes_avg}) with 
uncertainties included. This is expected also to contribute to a 
reduction in the currents in the \citet{2009ApJ...700L..88W}
solution. For these reasons the solution was presented as a proof 
of concept of the self-consistency procedure, rather than as a 
realistic attempt to model AR 10953.

\section{Modeling of AR 10953 including uncertainties\label{sec3}}

\subsection{Data\label{sec31}}

The calculations presented here use a set of data for NOAA AR 10953 
derived from the same Hinode/SOT SpectroPolarimeter (Tsuneta et al.\ 
2008) observations at 22:30~UT on 30 April 2007 previously used in 
\citet{2009ApJ...696.1780D} and \citet{2009ApJ...700L..88W}. 
The original Hinode data are again inverted
with the HAO/ASP inversion code developed by Paul Seagraves
at NCAR/HAO \citep{BWLcom} which was used previously. 
The code provides estimates of the uncertainties in each magnetic 
component derived from the $\chi^2$ variation of the least-squares fit 
results for the observed Stokes spectra. These error estimates may be
considered to provide a lower limit for the uncertainty in any single
measurement. Standard error propagation is applied to the 
transformations used to obtain the field components $(B_x,B_y,B_z)$ 
in the Hinode/SP field of view to give corresponding component
uncertainties $(\sigma_{B_x},\sigma_{B_y},\sigma_{B_z})$ 
\citep{1999SoPh..188....3L}.

As with the \citet{2009ApJ...696.1780D} approach, co-aligned MDI 
line-of-sight data are used initially to produce a potential field with 
a very large field of view, including AR 10954 roughly 15$^\circ$ to 
the east of AR 10953. However, the data-merging procedure deviates 
henceforth. The two datasets are re-sampled onto a $0.5\arcsec$ grid
(as compared to the original $0.32\arcsec$ and $1.8\arcsec$ resolution 
of the Hinode and MDI data, respectively). The computed MDI-derived 
image-plane potential vector field is merged with the coaligned 
Hinode/SP image-plane vector field (each component separately), by 
including a Hanning-function weighting at the boundary between the two 
datasets. This approach ensures a smooth transition between the data
over a few pixels and avoids introducing spurious vertical currents
due to sharp gradients between the two datasets. 

The resulting combined Hinode/MDI vector field data and the 
uncertainties for each component are cropped to a field of view similar 
in size to that used by \citet{2009ApJ...696.1780D} and 
\citet{2009ApJ...700L..88W}, and are re-sampled to the same 
$0.8\arcsec$ resolution used in the previous investigations, resulting
in a final $313\times 313$ grid. The $B_z$ values used here are flux 
balanced to within $1\%$. 

Values of $\alpha_0$ are calculated
for all points with $|B_z|>0.01\times \mbox{max}(B_z)$ using 
Equation~(\ref{eq:vmag_alpha}), with centered differencing used to 
approximate the derivatives. Uncertainties $\sigma_0$ for
the force-free parameter boundary values are calculated at each point 
from $(B_x,B_y,B_z)$ and $(\sigma_{B_x},\sigma_{B_y},\sigma_{B_z})$ 
using propagation of errors applied to the differencing scheme 
\citep{1999SoPh..188....3L}. 
A nominal large uncertainty value, equal to the maximum over all 
points, is assigned to points in the MDI field region. Points where 
the calculated $\alpha_0$ is zero because 
$|B_z|\leq 0.01\times\mbox{max}(B_z)$ are also assigned this nominal 
large uncertainty. The effect of this is that poorly-determined boundary 
values of $\alpha$ carry little weight in the decisions on new 
boundary values made by the self-consistency procedure using 
Equation~(\ref{eq:bayes_avg}).

\begin{figure}
\plotone{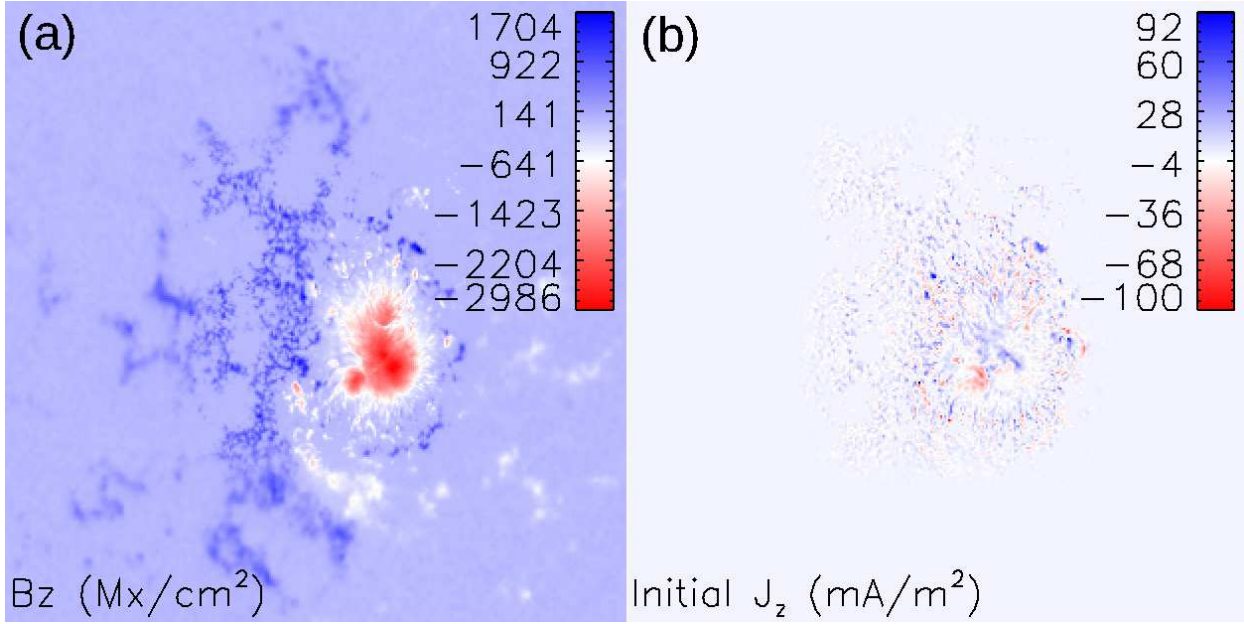}
\caption{The vector magnetogram boundary data for NOAA AR 10953 used 
in the magnetic field reconstructions in this paper. Panel (a): 
photospheric values of the vertical field $B_z$. Panel (b):  
initial photospheric values of the vertical electric current density 
$J_z$.
\label{fig:NewBCs}}
\end{figure}

Figure~\ref{fig:NewBCs} illustrates the resulting vector magnetogram
for NOAA AR 10953. Panel (a) shows the values of $B_z$, and panel (b)
shows the boundary values of $J_z$. The same color table 
(red-white-blue for negative-zero-positive values) used in 
Figure~\ref{fig:WheReg09Stereo} is adopted. Panel (b) 
may be compared with Figure~3 in \citet{2009ApJ...696.1780D} and 
Figure~3 in \citet{2009ApJ...700L..88W}, which show the current 
density values for the vector magnetogram used in the earlier studies. 
The current structures in the two sets of boundary data are very 
similar, including strong regions of current in the leading negative 
polarity spot, and less concentrated current structures in the more 
diffuse region of positive polarity. The new current map lacks the 
artifacts associated with the edge of the Hinode data region visible 
in the earlier data.

\begin{figure}
\plotone{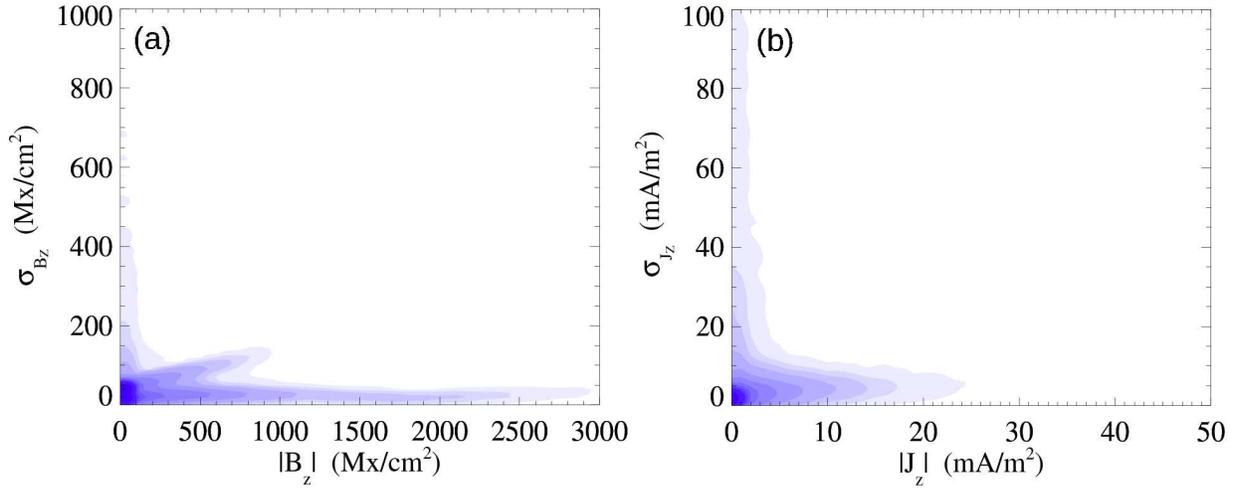}
\caption{The uncertainties in the vector magnetogram boundary values
of the field, represented using non-parametric estimates
of the density distributions of boundary points.  Panel (a): density 
distribution over values of $|B_z|$ and $\sigma_{B_z}$. Panel (b): 
density distribution over $|J_z|$ and $\sigma_{J_z}$.
\label{fig:SigNPdist}}
\end{figure}

Figure~\ref{fig:SigNPdist} illustrates
the quantitative values of the uncertainty estimates for $B_z$ and 
$J_z$ obtained for the vector magnetogram. This figure displays 
non-parametric estimates 
\citep[e.g.][]{SilvermanDensEst1986} of the density distribution of 
boundary points over values of $|B_z|$ and $\sigma_{B_z}$ [panel (a)],
and over values of $|J_z|$ and $\sigma_{J_z}$ [panel (b)]. The 
uncertainty values for $B_z$ illustrated in the panel (a) of 
Figure~\ref{fig:SigNPdist} are relatively small overall because the 
Hinode/SP-derived uncertainties are based solely on $\chi^2$ fitting 
values, and do not include estimates of detection levels or systematic 
errors which would increase the uncertainty. Hence the estimates presented
here should be considered to be lower bounds. The horizontal 
band of low uncertainty values for $B_z$ in panel (a) of the figure
represents the uniform noise in the MDI-derived areas. 
Panel (b) of Figure~\ref{fig:SigNPdist} shows a substantial horizontal 
band of large values of $|J_z|$ with relatively small uncertainties, 
representing significant currents in the boundary data, as well as a 
vertical band of small $|J_z|$ values with large uncertainties, 
representing points where the current is poorly determined.

\subsection{Results\label{sec32}}

The current-field iteration method \citep{2007SoPh..245..251W} is 
applied to the data, as outlined in section~\ref{sec23}, for 10 
self-consistency cycles with 30 Grad-Rubin iterations used at each 
cycle to construct $P$ and $N$ solutions. Each solution is calculated 
on a $313\times 313\times 300$ grid. The horizontal grid size matches
the boundary data, and the vertical grid size is chosen to include all
field lines which do not cross the sides of the computational volume.

The treatment of `missing information' associated with the absence of 
boundary conditions at the sides and top of the computational volume
is important for coronal magnetic field reconstruction. The 
\citet{2007SoPh..245..251W} method addresses this problem by neglecting 
currents on field lines which cross these boundaries. At each iteration 
for each solution, points ${\bf r}=(x,y,z)$ on the grid threaded by 
field lines which cross the sides or top boundaries have 
$\alpha ({\bf r})$ set to zero. This includes points $(x,y,0)$ on the 
grid in the lower boundary, 
so this procedure modifies the boundary values of the force-free 
parameter during the Grad-Rubin iterations. For boundary points 
subject to this modification the corresponding uncertainty in the 
force-free parameter is set equal to the nominal large uncertainty 
value discussed in section~\ref{sec31}. This choice minimizes the 
effect of the zero values during the construction [according to 
Equation~(\ref{eq:bayes_avg})] of a new set of $\alpha$ boundary values 
at the end of a self-consistency cycle. The choice of the vertical grid
size mentioned above also reduces the effect of this modification of
the boundary conditions.

\begin{figure}
\epsscale{0.80}
\plotone{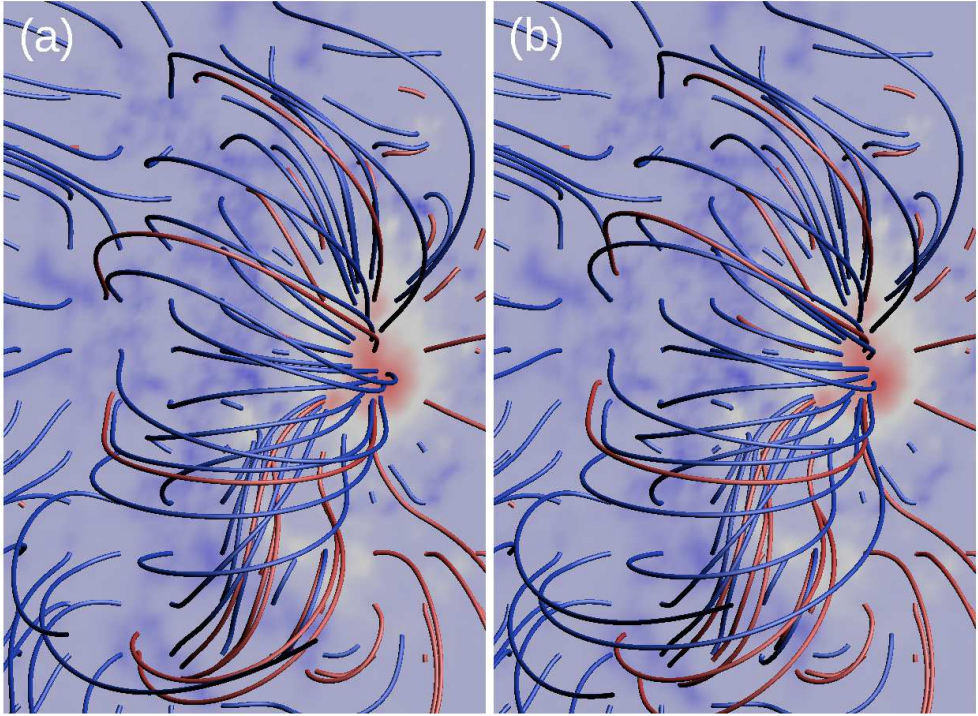}
\caption{The self-consistent solution for the coronal magnetic field
of NOAA AR 10953 obtained using the new vector magnetogram boundary 
data including uncertainties. The presentation is the same as for 
Figure~\ref{fig:WheReg09Stereo}. Panel (a) shows the $P$ solution 
and panel (b) shows the $N$ solution, after 10 self-consistency 
cycles. The color images in the background are the boundary values of 
$B_z$, and the blue/red curves are field lines originating in the 
$P$/$N$ polarities. The sets of field lines for the $P$ solution 
in panel (a) and for the $N$ solution in panel (b) agree closely.
\label{fig:NewSolStereo}}
\end{figure}

The self-consistency procedure converges in less than ten cycles. 
Figure~\ref{fig:NewSolStereo} shows the self-consistent $P$ and
$N$ solutions [panels (a) and (b) respectively] obtained after 
10 cycles, using the same presentation as for the 
\citet{2009ApJ...700L..88W} solution shown in 
Figure~\ref{fig:WheReg09Stereo}. 
Once again the red-white-blue images in the background show the 
boundary values of $B_z$, and the blue and red curves are field lines 
originating in $P$ and $N$ polarities respectively. The sets of field 
lines shown in panels (a) and (b) agree closely, demonstrating
the self-consistency achieved. Some specific differences are visible for
field lines extending higher into the computational volume. 
Table~\ref{tbl-1} lists values of the vector field comparison metrics
introduced by \citet{2006SoPh..235..161S} in the context of testing
nonlinear force-free methods on known solutions. The more stringent of 
these checks, the mean vector error and the normalized vector error, 
are both about $2\%$ for the $N$ solution compared with the $P$ 
solution, which are small values for these metrics in this context
\citep{2006SoPh..235..161S,2008SoPh..247..269M,2010A&A...516A.107W}. 
The  energies of the $P$ and $N$ solutions shown in 
Figure~\ref{fig:NewSolStereo} are $E/E_0\approx 1.0822$ and 
$E/E_0\approx 1.0819$, which differ by less 
than $0.03\%$. The two fields are sufficiently similar that we refer to 
the result as representing a single self-consistent solution.

The solution shown in 
Figure~\ref{fig:NewSolStereo} is considerably more distorted from 
the potential configuration than the \citet{2009ApJ...700L..88W} 
solution shown in Figure~\ref{fig:WheReg09Stereo}, consistent with the 
larger magnetic free energy (as discussed in section~\ref{sec24}, the 
earlier solution has energy $E/E_0\approx 1.02$). The energy of the 
self-consistent solution obtained here ($E/E_0\approx 1.08$) is in the 
middle of the range of values reported in \citet{2009ApJ...696.1780D} 
for different nonlinear force-free methods applied to preprocessed 
boundary data (the values were in the range 
$E/E_0\approx 1.03$--$1.25$). The energy of the self-consistent 
solution obtained here is larger than the energy $E/E_0\approx 1.03$ 
of the $P$ solution constructed at the first cycle from the 
vector magnetogram boundary values, but is smaller than the energy
$E/E_0\approx 1.17$  of the $N$ solution constructed initially. 
The averaging of $\alpha$ values between foot points of field lines 
has led to a solution with an intermediate energy. The potential field 
has an energy $E_0\approx 8.40\times 10^{32}\,{\rm erg}$, so the 
magnetic free energy for the region is about 
$7\times 10^{31}\,{\rm erg}$.

\begin{figure}
\plotone{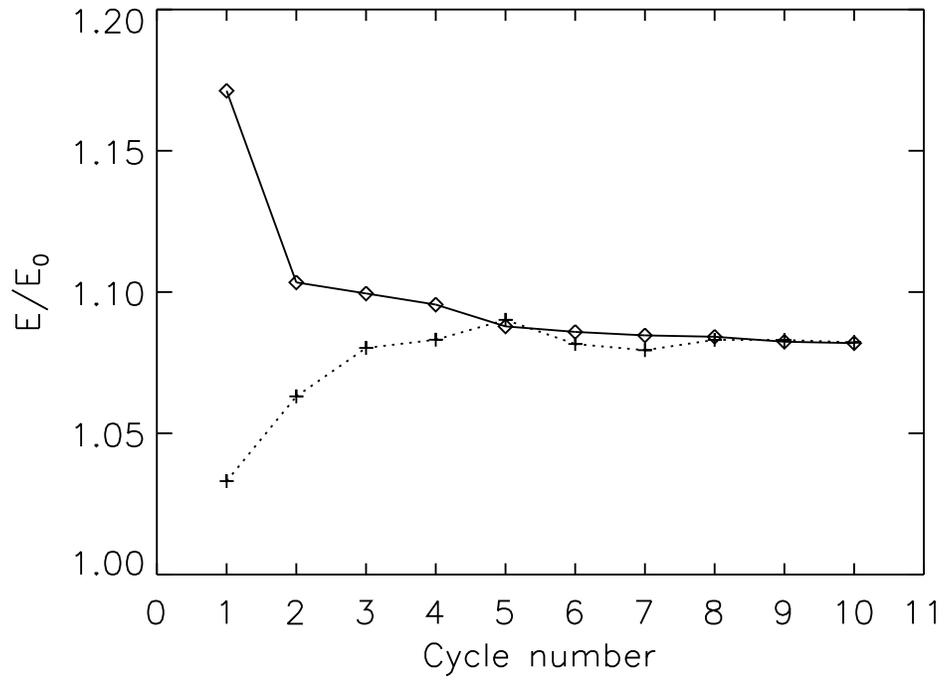}
\caption{The energy of the $P$ solutions (plus signs) and the $N$
solutions (diamonds) constructed at each self-consistency cycle, in
units of the reference potential field energy.
\label{fig:SCEnergyVsIt}}
\end{figure}

Figure~\ref{fig:SCEnergyVsIt} illustrates the convergence of the 
self-consistency procedure, showing the energies of the $P$ and $N$ 
solutions obtained at the end of each self-consistency cycle, versus 
cycle number. The $P$-solution energies are indicated by plus signs, 
and the $N$-solution energies by diamonds. The energy of the first $N$ 
solution is substantially larger than the energy of the first $P$ 
solution, as mentioned above. The solutions constructed at 
intermediate cycles approach one another in energy, converging in 
about eight cycles. During the initial self-consistency cycles, the 
Grad-Rubin iterations exhibit oscillation, as discussed in
section~\ref{sec22}. The boundary data contains large values of 
the electric current density which prevent strict convergence of the
Grad-Rubin method. This introduces some arbitrariness into the 
modeling, because the exact $P$ and $N$ solutions, and hence the 
energies shown in Figure~\ref{fig:SCEnergyVsIt} depend on the choice 
of a stopping iteration. This problem is discussed in more detail 
below. The Grad-Rubin solutions constructed for later cycles converge 
quite strictly (a fixed point of the iteration scheme is achieved). 

\begin{figure}
\plotone{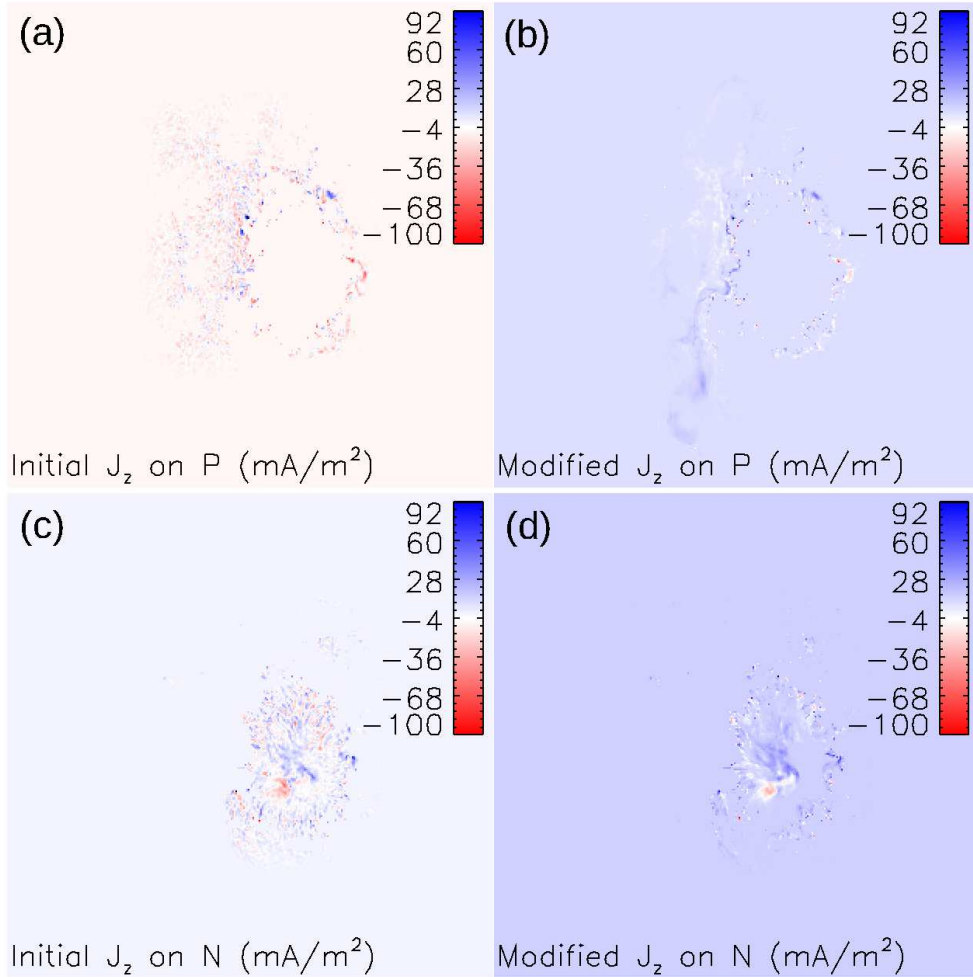}
\caption{Initial boundary conditions on $J_z$ and the boundary 
conditions on $J_z$ for the self-consistent solution, for each polarity.
Panel (a) shows the initial $J_z$ values over $P$, and panel (b)
shows the self-consistent 
$J_z$ values over $P$. Panels (c) and (d) show the same quantities,
respectively, over $N$.
\label{fig:JzSCPN}}
\end{figure}

\begin{figure}
\plotone{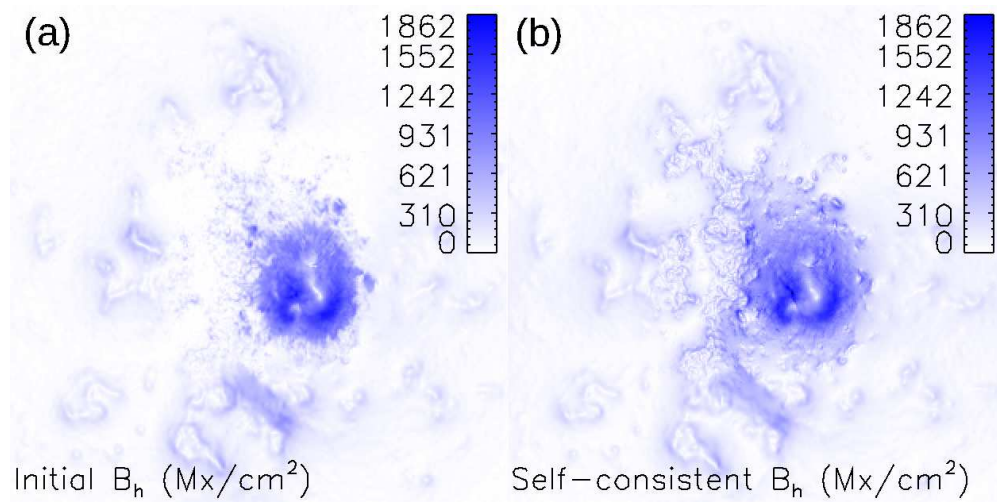}
\caption{Panel (a): The horizontal field at the photosphere for the 
vector magnetogram. Panel (b): the horizontal field at the 
photosphere for the self-consistent solution.
\label{fig:BhChange}}
\end{figure}

The self-consistency procedure changes the boundary values of
the vertical electric current density and hence the boundary values of
the horizontal field ${\bf B}_h=(B_x,B_y)$, but does not change 
the boundary values of $B_z$. Figure~\ref{fig:JzSCPN} illustrates the 
changes in the vertical current density $J_z$ over the $P$ and $N$ 
polarities in the boundary. The self-consistency procedure reduces 
the largest values of $|J_z|$ over both polarities, but retains some 
large currents. Many of the structures in the distribution of $J_z$ 
observed in the initial data have survived, and in particular the 
patterns of strong-current in the leading negative polarity spot are 
preserved [see panels (c) and (d) of Figure~\ref{fig:JzSCPN}], 
although the current 
distribution has changed in detail. Figure~\ref{fig:BhChange} 
illustrates the corresponding boundary values of the magnitude of the 
horizontal field at the photosphere, for the initial vector magnetogram 
and for the final self-consistent solution. 
Panel (a) of Figure~\ref{fig:BhChange} shows the horizontal field 
for the magnetogram data,
$B_h^i=\left[\left(B_x^i\right)^2+\left(B_y^i\right)^2\right]^{1/2}$, 
and panel (b) shows the the horizontal field for the 
self-consistent solution,
$B_h^f=\left[\left(B_x^f\right)^2+\left(B_y^f\right)^2\right]^{1/2}$. 
The maximum value of the two fields is very similar, and there is a good
qualitative correspondence between features in the two images, although
the details have changed. 

\begin{figure}
\plotone{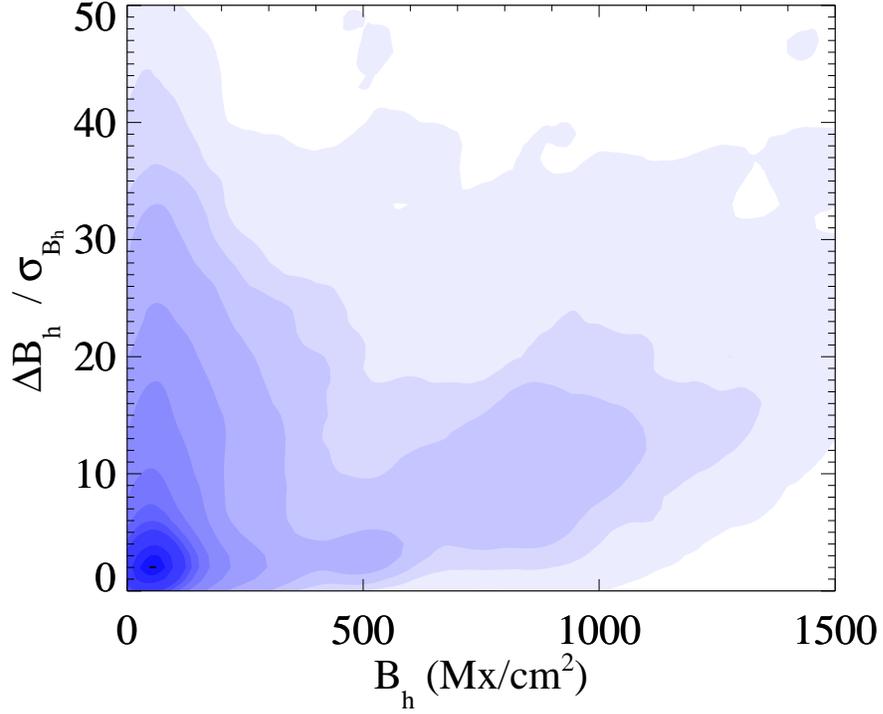}
 \caption{Quantitative
changes in the horizontal field at the photosphere produced by the
self-consistency procedure. A non-parametric estimate of the density
distribution of boundary points, showing  $\Delta B_h/\sigma_{B_h}$ as a
function of the initial ${B_h}$ magnitude, where $\sigma_{B_h}$ is the
uncertainty in the magnitude of the initial horizontal field, and
$\Delta B_h=\left[(B_x^f-B_x^i)^2+(B_y^f-B_y^i)^2\right]^{1/2}$ is the
change in the horizontal field, with $i$ denoting values in the initial
vector magnetogram, and $f$ the final values in the self-consistent
solution. \label{fig:SigBhNPdist}}
\end{figure}

Figure~\ref{fig:SigBhNPdist} illustrates the
quantitative changes in the horizontal vector field, showing a
non-parametric estimate of the density of points (cf.\
Figure~\ref{fig:SigNPdist}) over values of $\Delta B_h/\sigma_{B_h}$ as
a function of the measured ${B_h}$, where $\Delta
B_h=\left[(B_x^f-B_x^i)^2+(B_y^f-B_y^i)^2\right]^{1/2}$ and $\sigma_{B_h}$
is the uncertainty in $B_h^i$. Only points in the Hinode data region are
included in this comparison. This figure indicates that the vast
majority of points undergo small changes, but there is a substantial
low-density tail of points subject to larger changes. The average
absolute change in the horizontal field is 
$\langle \Delta B_h\rangle\approx 170\,{\rm Mx}/{\rm cm}^2$, and the 
average ratio of the changes to the uncertainties is 
$\langle \Delta B_h/\sigma_{B_h}\rangle \approx 9$. Many points 
are subject to changes in the horizontal field substantially larger 
than the corresponding 
uncertainty. As discussed in section~\ref{sec31}, the uncertainties 
assigned to the vector magnetogram data are lower bounds. Also, the 
initial boundary data are physically inconsistent with the force-free 
model (the departure from the model is not due only to observational
uncertainties). The absolute 
changes in the horizontal field are comparable to those produced by 
preprocessing \citep{2009ApJ...696.1780D}. 

A measure of the inconsistency between the boundary data and the 
force-free model is given by the estimates of the net forces and 
torques on the field in $z>0$ which may be calculated from the 
boundary data 
\citep{1969SvA....12..585M,1984ApJ...283..349A,1989SoPh..120...19A},
as discussed in section~\ref{sec22}. Table~\ref{tbl-2} presents the 
values of the forces and torques for the initial vector magnetogram 
data, and for the self-consistent $P$ and $N$ solutions. The components
of the initial force are of order a few percent, and the components of
the initial torque are of 
order $10\%$, in terms of relevant characteristic values. For the 
self-consistent solutions the force components are reduced to 
$< 0.1\%$, and the components of the torques are reduced to 
$\lesssim 0.1\%$ (in most cases much less), with the same scaling. 
The changes in the horizontal field and hence current illustrated in 
Figures~\ref{fig:JzSCPN}--\ref{fig:SigBhNPdist} are necessary to 
achieve this consistency with the force-free model.

As mentioned above, the Grad-Rubin iterations do not converge exactly
during the early and intermediate self-consistency cycles, and
the choice of the final iteration for each cycle introduces some
arbitrariness into the procedure. To investigate this effect the
calculation is repeated, with $N_{\rm GR}=20$ Grad-Rubin iterations per
cycle for both the $P$ and $N$ solutions (rather than $N_{\rm GR}=30$), 
also with $N_{\rm GR}=40$ iterations per cycle. The Grad-Rubin 
iterations converge in $\lesssim 20$ cycles for later cycles, so 
$N_{\rm GR}=20$ is approximately the minimum suitable choice. In each 
case, the self-consistency procedure is found to converge in less than 
10 cycles. The self-consistent fields obtained with $N_{\rm GR}=20$ 
and $N_{\rm GR}=40$ are qualitative very similar to the $N_{\rm GR}=30$
result, i.e.\ the field lines are visually very similar. The energies
are also remarkably similar, with all three solutions having 
$E/E_0\approx 1.08$, to the stated digits. Table~\ref{tbl-3} quotes 
the vector field comparison metrics from \citet{2006SoPh..235..161S},
comparing the $P$ solutions of the two new cases 
($N_{\rm GR}=20$ and $N_{\rm GR}=40$) with the original case 
($N_{\rm GR}=30$). The values are substantially larger than for the
comparison of the individual $P$ and $N$ solutions 
(e.g.\ Table~\ref{tbl-1}), but are smaller than the typical differences 
introduced by varying treatments of the unknown side and top boundary
conditions in the application of nonlinear force-free methods to known
solutions \citep{2006SoPh..235..161S,2008SoPh..247..269M}. Hence in the
context of nonlinear force-free modeling, the solutions may be 
considered very similar.

Figure~\ref{fig:VaryGRIT} 
illustrates the boundary distributions of $J_z$ for the three 
self-consistent $N$ solutions. Panel (a) shows the signal to noise 
ratio values $\alpha_0/\sigma_0$ over the vector magnetogram, which
determine how well the initial boundary conditions on $J_z$ are 
preserved at different locations in the boundary by the 
self-consistency procedure, according to the first of 
Equations~(\ref{eq:bayes_avg}). Panels (b), (c), and (d) show the 
boundary conditions on $J_z$ in the self-conistent solutions for the 
cases $N_{\rm GR}=20$, $N_{\rm GR}=30$, and $N_{\rm GR}=40$ 
respectively, using the same presentation as Figures~\ref{fig:NewBCs} 
and~\ref{fig:JzSCPN}. Panels (b)--(d) are visually very similar. 
The same structures are observed, and the range of values of $J_z$ is 
very similar. Comparison with the signal to noise ratio map [panel (a)]
illustrates qualitatively how the current is unchanged at locations 
where $\alpha_0$ (or $J_z$) is well-determined. These results suggest 
that the self-consistency procedure, including uncertainties, is 
identifying a particular force-free solution independent of this element 
of arbitrariness.

\section{Discussion\label{sec4}}

A solution to the nonlinear force-free model is constructed for NOAA 
solar active region AR 10953, based on a vector magnetogram derived from
Hinode/SOT observations at 22:30~UT on 30 April 2007. The solution 
applies the `self-consistency' procedure of \citet{2009ApJ...700L..88W},
for the first time taking into account uncertainties in the boundary 
data. The boundary data are subject to improved techniques for data
merging by comparison with earlier studies using the same Hinode/SOT
observations \citep{2009ApJ...696.1780D,2009ApJ...700L..88W}.
The self-consistency procedure addresses the problem that
vector magnetogram data are inconsistent with the force-free model
\citep{2009ApJ...696.1780D,WheEtAl2010}. In particular, the solar 
boundary data provide two possible force-free solutions (the $P$ and 
$N$ solutions, corresponding to choosing boundary conditions for 
the vertical electric current density $J_z$ from the positive or the 
negative polarities of the boundary field respectively), which
are generally found to be significantly different. The $P$ and the $N$
soutions are distinctly different for the vector magnetogram data used
here. The self-consistency procedure modifies the boundary conditions 
on $J_z$ during a sequence of cycles in which $P$ and $N$ solutions 
are constructed using Grad-Rubin iteration \citep{2007SoPh..245..251W}, 
and arrives at a solution to the force-free model with boundary values 
close to, but not exactly matching, the observational vector 
magnetogram data. When uncertainties in the boundary data are taken 
into account, the boundary values of $J_z$ are preserved more closely 
at points having smaller uncertainties. 

Self-consistency modeling has 
two advantages over conventional force-free modeling. First, force-free 
methods tend to fail to strictly converge when applied to solar data, 
and hence do not accurately solve the force-free model. The 
self-consistency 
procedure achieves strict convergence. Second, the method identifies 
a single solution, rather than two solutions. Active region AR 10953 
was previously modelled using the self-consistency procedure in 
\citet{2009ApJ...700L..88W}, but in that case uncertainties were not 
included, and the results were regarded as a proof of concept of the 
method, rather than a full demonstration of its capabilities. This 
paper presents a more comprehensive demonstration and investigation.

\begin{figure}
\epsscale{0.8}
\plotone{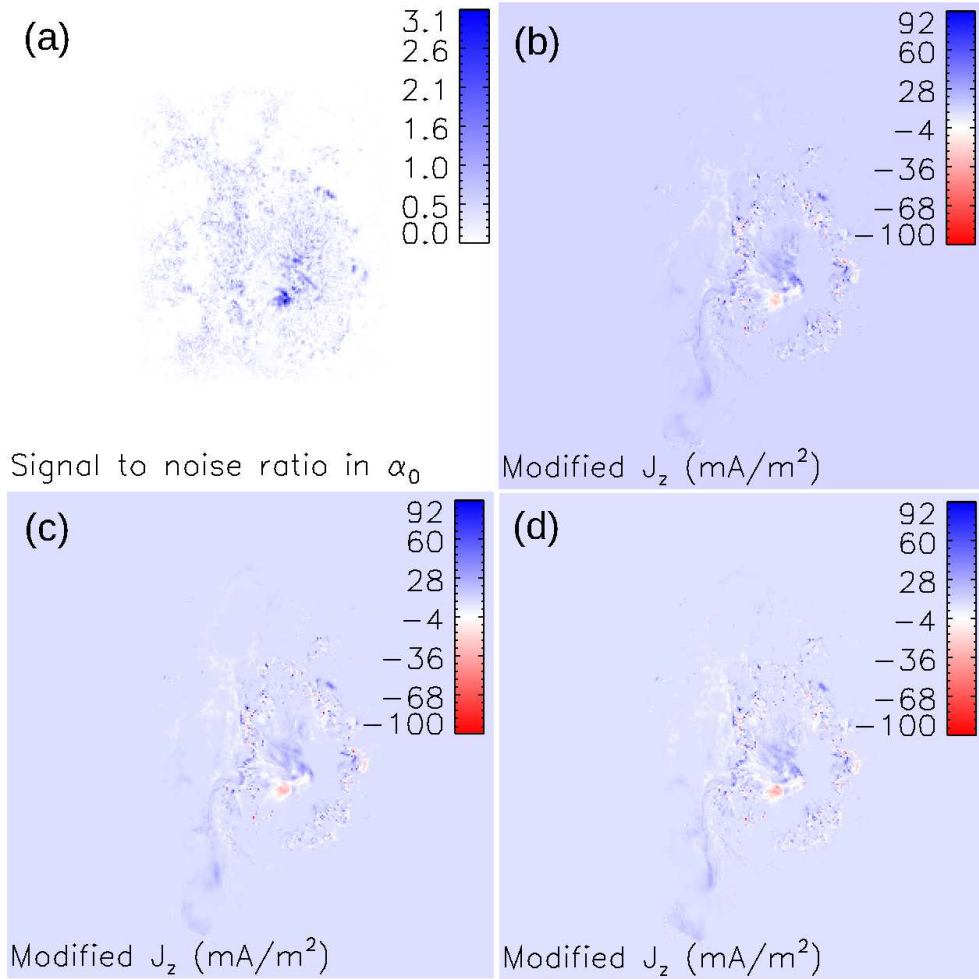}
\caption{Panel (a) shows a map of the signal to noise ratio for 
$\alpha_0$, which determines how well the initial boundary conditions
on $J_z$ are preserved by the self-consistency procedure. Panels
(b), (c), and (d) show the boundary conditions on $J_z$ for the 
self-consistent $N$ solutions obtained using 20, 30, and 40 Grad-Rubin 
iterations per self-consistency cycle respectively.
\label{fig:VaryGRIT}}
\end{figure}

The self-consistent solution obtained for active region AR 10953 is 
substantially non-potential, with a coronal magnetic energy 
$E/E_0\approx 1.08$, where $E_0$ is the energy of the potential 
field with the same vertical component of the magnetic field in the 
boundary. This energy is significantly larger than the energy 
$E/E_0\approx 1.02$ of the solution obtained in 
\citet{2009ApJ...700L..88W}, which is attributed to the neglect of 
uncertainties in the earlier paper. The non-potentiality
of the region is due to strong currents in the inner parts of the 
region, in particular associated with the southern part of the the 
negative polarity leading spot, as shown in Figure~3. These currents 
were reduced in the earlier solution by the self-consistency averaging
procedure [Equations~(\ref{eq:bayes_avg})] due to the omission of 
uncertainties (or rather, the treatment of all
boundary points as having equal uncertainty). With uncertainties
included correctly, these currents are preserved because they are
in regions where the field, and hence current, is well-determined,
as shown in Figure~\ref{fig:VaryGRIT}.

The energy of the new self-consistent solution is intermediate between
the energies of the $P$ and $N$ solutions constructed from the vector 
magnetogram data ($E/E_0\approx 1.03$ for the $P$ solution, and 
$E/E_0\approx 1.17$ for the $N$ solution), and is also in the middle of
the range of energies found by \citet{2009ApJ...696.1780D} using 
force-free codes applied to boundary data derived from the same 
Hinode/SOT observations but subject to the preprocessing procedure 
\citep{2006SoPh..233..215W,2008SoPh..247..249W}, which is not used 
here. Recently \citet{2010ApJ...715.1566C} presented a more detailed
account of force-free modeling of AR 10953 with two Grad-Rubin codes
applied to the $N$-polarity
\citet{2009ApJ...696.1780D} boundary data, and reported energies 
$E/E_0\approx 1.27$ and $E/E_0\approx 1.31$. The energies obtained 
by these authors are different from the energy of the initial $N$ 
solution obtained here for a number of reasons. The Hinode/MDI boundary 
data used here is prepared differently, as discussed in 
section~\ref{sec31}, 
and preprocessing has not been applied. The lack of convergence of 
the initial $P$ and $N$ solutions obtained here means that a range of 
energies is possible, depending on the choice of stoppping iteration
(in practise we find energies for the $N$ solution in the range 
$E/E_0\approx 1.16$ to $E/E_0\approx 1.22$). The present method 
treats the side and top boundaries in a different way to the 
\citet{2010ApJ...715.1566C} method, and in particular we neglect 
currents on field lines which cross the side and top boundaries, which 
tends to reduce the energy. Finally, the stated 
\citet{2010ApJ...715.1566C} and \citet{2009ApJ...696.1780D} energies 
refer to a smaller sub-region of the computational domain. Because of 
these differences we do not attempt a more detailed comparison with 
the earlier results.

A degree of arbitrariness is introduced into the modeling by the 
lack of strict convergence of the Grad-Rubin iterations used to 
construct the $P$ and $N$ solutions at early and intermediate 
self-consistency cycles. The results of the self-consistency procedure 
may depend on the choice of the number of Grad-Rubin iterations. This 
effect is investigated by repeating the calculation with 
$N_{\rm GR}=20$ and $N_{\rm GR}=40$ iterations at each cycle (rather 
than $N_{\rm GR}=30$). The two new self-consistent solutions are 
remarkably similar to the first solution, and in particular all three 
solutions have energy $E/E_0\approx 1.08$. This suggests that 
self-consistency modeling provides a solution to the problem of 
reliable estimation of the coronal magnetic free energy of an active 
region, once uncertainties in the boundary data are incorporated. The 
inclusion of uncertainties preserves boundary conditions on current at 
points in the boundary where the currents are most accurately 
determined. The three self-consistent solutions obtained with different 
numbers of Grad-Rubin iterations have very similar boundary 
distributions of $J_z$. 

The self-consistency procedure (including uncertainties) is a 
promising candidate for routine reconstruction of coronal magnetic 
fields. The technique is shown to produce a highly non-potential 
and accurately force-free coronal field model for NOAA AR 10953 from 
Hinode/SOT vector magnetogram data. The estimate of the free energy of
the magnetic field appears to be robust based on the results obtained
with varying numbers of Grad-Rubin iterations per self-consistency
cycle. The availability of high-quality vector magnetogram data over 
the next solar cycle from Hinode/SOT and SDO/HMI will provide ample 
opportunities for further testing, development, and application of 
the method, to investigate many questions in the physics of solar 
activity.

\acknowledgments
Hinode is a Japanese mission developed and launched by ISAS/JAXA,
collaborating with NAOJ as a domestic partner, and NASA and STFC (UK) as
international partners. Scientific operation of the Hinode mission is
conducted by the Hinode science team organized at ISAS/JAXA. This team
mainly consists of scientists from institutes in the partner countries.
Support for the post-launch operation is provided by JAXA and NAOJ
(Japan), STFC (U.K.), NASA, ESA, and NSC (Norway). We extend our thanks
to Bruce Lites for providing the full inversion results for the 
data used herein and Paul Seagraves for developing the code at NCAR/HAO.  
KDL appreciates funding from NSF SHINE grant ATM-0454610 and NPDE code
from Graham Barnes. Mike Wheatland acknowledges support from the same 
NSF SHINE grant during a collaborative visit to NWRA in Boulder.

{\it Facilities:} \facility{Hinode}, \facility{SOHO}.

\clearpage

\clearpage

\clearpage

\begin{table}
\begin{center}
\caption{Vector field comparison metrics for the calculated 
self-consistent $P$ and $N$ solution fields, for the default 
calculation (with 30 Grad-Rubin iterations per self-consistency cycle).
\label{tbl-1}}
\vspace{0.2cm}
\begin{tabular}{cc}
\tableline\tableline
Metric\tablenotemark{a} & Value \\
\tableline
$1-\mbox{CS}$ & $8\times 10^{-6}$  \\
$1-\mbox{VC}$ & $3\times 10^{-4}$  \\
MVE & $2\times 10^{-2}$ \\
NVE & $2\times 10^{-2}$ \\
\tableline
\end{tabular}


\tablenotetext{a}{These metrics were introduced 
by \citet{2006SoPh..235..161S}. CS is the Cauchy-Schartz metric, 
$\mbox{CS}=M^{-1}\sum_i {\bf B}_{1i}\cdot {\bf B}_{2i}
  /(B_{1i}B_{2i})$, where ${\bf B}_1$ is the $P$ solution
and ${\bf B}_2$ is the $N$ solution, the sum is over all 
grid points enumerated by $i$, and $M=313\times 313\times 300$ is 
the total number of grid points. VC is the vector corrrelation, 
$\mbox{VC}=\sum_i{\bf B}_{1i}\cdot {\bf B}_{2i}\left/
  \sqrt{\sum_i B_{1i}^2\sum_i B_{2i}^2}\right.$. MVE is the mean 
vector error,
$\mbox{MVE}=M^{-1}\sum_i|{\bf B}_{1i}-{\bf B}_{1i}|/B_{1i}$, and
NVE is the normalized vector error, 
$\mbox{NVE}=\sum_i|{\bf B}_{1i}-{\bf B}_{2i}|
  \left/\sum_iB_{1i}\right.$.}
\end{center}
\end{table}

\begin{table}
\begin{center}
\caption{Estimates of the components of the net force ${\bf F}$ and 
  net torque ${\bf \Gamma}$ on the coronal volume above NOAA AR 
  10953.\label{tbl-2}}
\vspace{0.2cm}
\begin{tabular}{cccc}
\tableline\tableline
Quantity & Vector magnetogram & Self-consistent P solution &
  Self-consistent N solution\\
\tableline
$F_x/F_0$\tablenotemark{a} & $-1.2\times 10^{-2}$ 
  & $4.8\times 10^{-4}$ & $4.3\times 10^{-4}$\\
$F_y/F_0$ & $3.9\times 10^{-2}$ & $7.1\times 10^{-4}$ 
  & $6.5\times 10^{-4}$ \\
$F_z/F_0$ & $-7.0\times 10^{-2}$ & $-5.1\times 10^{-5}$ 
  & $7.1\times 10^{-4}$ \\
$\Gamma_x/\Gamma_0$\tablenotemark{b} & $-1.2\times 10^{-1}$ 
  & $1.1\times 10^{-3}$ & $1.9\times 10^{-4}$ \\
$\Gamma_y/\Gamma_0$ & $1.2\times 10^{-1}$ & $-2.5\times 10^{-4}$ 
  & $1.9\times 10^{-4}$ \\
$\Gamma_z/\Gamma_0$ & $3.9\times 10^{-2}$ & $-7.6\times 10^{-5}$
  & $-1.0\times 10^{-4}$\\
\tableline
\end{tabular}


\tablenotetext{a}{The forces are in units of $F_0=\int p_B \,dxdy$
with $p_B=(B_x^2+B_y^2+B_z^2)/2\mu_0$, where the integral is performed
over the magnetogram area with the origin for the coordinates at the 
lower left corner.}
\tablenotetext{b}{The torques are in units of $\Gamma_0=\int x p_B \,dxdy$,
with $p_B=(B_x^2+B_y^2+B_z^2)/2\mu_0$, where the integral is performed
over the magnetogram area with the origin for the coordinates at the 
lower left corner.}
\end{center}
\end{table}

\begin{table}
\begin{center}
\caption{Vector field comparison metrics for the
self-consistent $P$ solutions calculated with different numbers of 
Grad-Rubin iterations per self-consistency cycle. The comparisons are
for the solutions with $N_{\rm GR}=20$ and $N_{\rm GR}=30$, and for the
solutions with $N_{\rm GR}=20$ and $N_{\rm GR}=30$.
\label{tbl-3}}
\vspace{0.2cm}
\begin{tabular}{ccc}
\tableline\tableline
Metric\tablenotemark{a} & $N_{\rm GR}=30$ and $N_{\rm GR}=20$ & 
  $N_{\rm GR}=30$ and $N_{\rm GR}=40$ \\
$1-\mbox{CS}$ & $5\times 10^{-4}$ & $4\times 10^{-4}$ \\
$1-\mbox{VC}$ & $3\times 10^{-3}$ & $3\times 10^{-3}$ \\
MVE & $0.08$ & $0.09$ \\
NVE & $0.08$ & $0.09$ \\
\tableline
\end{tabular}


\tablenotetext{a}{The definitions of the metrics are the same as 
given in Table~\ref{tbl-1}, with ${\bf B}_1$ representing the 
$N_{\rm GR}=30$ solution and ${\bf B}_2$ the solution with either
$N_{\rm GR}=20$ or $N_{\rm GR}=30$.}
\end{center}
\end{table}

\end{document}